# Comparative study of the self-heating effect in the accumulation and inversion mode FinFETs

A.E. Atamuratov
*Physics department*
Urgench State University
Urgench, Uzbekistan
atabek.atamuratov@yahoo.com

B.O. Jabbarova
*Physics department*
Urgench State University
Urgench, Uzbekistan
bahorbahor1989@mail.ru

E.Sh. Xaitbayev
*Master student*
Urgench State University
Urgench, Uzbekistan
eldorkhaitbaev@gmail.com

D.R. Rajapov
*Master student*
Urgench State University
Urgench, Uzbekistan
dilshodbekrajapov5@gmail.com

M.M. Khalilloev
*Physics department*
Urgench State University
Urgench, Uzbekistan
x-mahkam@urdu.uz

*Abstract-* In this work, the self-heating effect in inversion mode FinFET and Junctionless (accumulation mode) FinFET is compared and the influence of the different electrical and geometric parameters on the self-heating effect is considered. It is shown, that the lattice temperature in the channel center is higher in junctionless (JL), accumulation mode FinFET than in inversion mode FinFET with the same parameters. The difference in the temperature is explained by the difference in the structure, particularly by the difference in the doping level of the channel for these transistors. The dependence of the self-heating effect on the doping depth and doping profile of the source and drain areas along the channel were considered.

## I. Introduction

The main trends of nanoelectronics are decreasing transistor sizes to provide decreasing energy consumption and increasing the integrity level of integral circuits (IC) [1,2,3]. Metal-oxide-semiconductor field effect transistors (MOSFET) are one of the main elements of the IC. Decreasing the MOSFET sizes leads to arising different types of degradation effects, particularly short channel effects [4, 5], and self-heating effects [6, 7, 8].

Short channel effects is significant for planar MOSFET at nanometer sizes. To increase immunity against short channel effects, fin field effect transistors (FinFET) [9] instead of planar MOSFET were suggested. FinFET works in inversion mode as well as MOSFET.

For simplification of the FinFET technology it was suggested new, junctionless FinFET transistor structure, which works in accumulation mode [10, 11] (Fig.1). In this new transistor technology thermic process which is used to form the source and drain areas by diffusion is absent. This thermic process in nanosized areas can lead to degradation of some parameters of the transistor. Therefore this technology is preferable and is more simple. Besides it, JL FinFET also shows high immunity against short-channel effects.

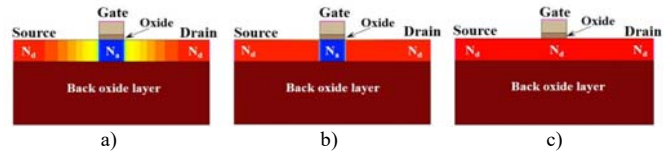

Fig. 1. The structures of the FinFET with analytical (a) and constant (b) doping profile in source and drain areas, and JL FinFET (c)

One of the features of the FinFET and JL FinFET is the fabrication of them on the basis of silicon-on-insulator (SOI) structures. It means the bottom of the channel of the transistors has contact with back oxide. Because the oxide materials have small thermal conductivity, in the channel the self-heating effect (SHE) can take place [12, 13]. The self-heating effect is an important problem in providing reliability of ICs. Therefore studying the self-heating effect in FinFET is very important problem. Separately this effect in FinFET and JL FinFET [6, 7, 8] was considered, however, comparison of this effect in such transistors with the same parameters is not carried out yet. In this work, the comparison of the self-heating effect in such transistors with the same parameters is considered.

## II. SIMULATION CONDITIONS AND TRANSISTOR PARAMETERS

FinFET differs from JL FinFET by the high doping level of the channel in JL FinFET and by the presence the source and drain areas which is doped with high concentrations in FinFET. Therefore it is expedience to investigate the influence of the parameters of these areas on SHE. In this work, it was simulated the dependence of the temperature in the center of the channel on the doping profile and on the depth of the doping along the channel in FinFET. It was considered the analytical and constant doping profile of the source and drain areas.

TCAD Sentaurus program was used in the simulation. The transistor parameters indicated in Table 1 were used. To account for thermic effects the thermodynamic transport model with quantum correction is used. In the mobility model, the doping dependence and velocity saturation at the high field were accounted. For calibration, the used physical models the I-V curves carried out in the simulation and experiment [14] were compared (Fig.2).

TABLE I. PARAMETERS OF THE FINFET AND JL FINFET USED IN THE SIMULATION

| Parameter | Designation | JLFinFET | FinFET |
|---|---|---|---|
| Dopin level in the channel | N | $5 \cdot 10^{18}$ cm$^{-3}$ (n-tip) | $10^{16}$ cm$^{-3}$ (p-tip) |
| Doping level in source and drain areas | $N_d$ | $5 \cdot 10^{18}$ cm$^{-3}$ (n-tip) | $5 \cdot 10^{18}$ cm$^{-3}$ (n-tip) |
| Gate oxide (HfO$_2$) layer thickness | $t_{ox}$ | 6,7 nm ($t_{eff}$ = 1,2 nm) | 6,7 nm ($t_{eff}$ = 1,2 nm) |
| Channel thickness | $T_{si}$ | 9 nm | 9 nm |
| Channel width | $W_b$ | 22 nm | 22 nm |
| Back oxide layer thickness | $T_{box}$ | 145 nm | 145 nm |
| Gate length | $L_{gate}$ | 10 nm | 10 nm |

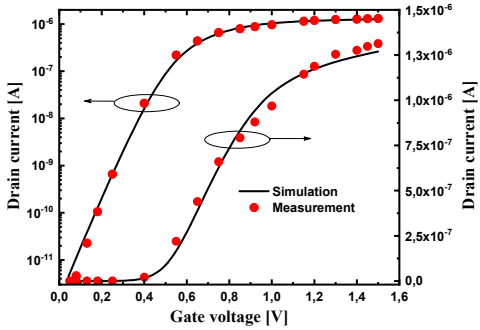

Fig.2. Comparing of the Id-Vg characteristics for SOI JL FinFET carried out in simulation and experiment.

## III. RESULTS OF SIMULATIONS AND DISCUSSION

The doping level of the channels for the considered JL FinFET and FinFET is $5 \cdot 10^{18}$ cm$^{-3}$ va $10^{16}$ cm$^{-3}$ respectively, which is typical for such transistors. The maximal Doping level at an analytical profile of doping the source and drain areas of FinFET is $5 \cdot 10^{18}$ cm$^{-3}$. The results of the simulation of Id-Vg dependence for FinFET as well as for JL FinFET is shown in Fig.3.

It is seen in the figure, in the saturation region a drain current in JL FinFET is higher than in FinFET. It results in a higher lattice temperature in the channel center of JL FinFET (Fig.4). The resulting temperature in the channel is defined by the heat generation as well as by the heat dissipation rates. The heat generation rate depends on the drain current, while the heat dissipation rate depends on the thermal properties of the materials surrounding the channel. In our consideration, the materials and geometry surrounding the channel is the same for both transistors and therefore the heat dissipation rate is also should be the same. Therefore in our case difference in the channel lattice temperature for considered transistors is defined only by the difference in the heat generation rate which is defined by the difference in a drain current in the transistors.

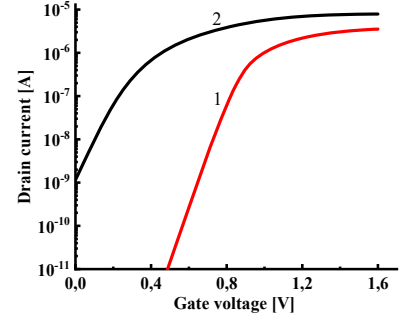

Fig.3. Id-Vg characteristics of inversion mode FinFET (1) and accumulation mode JL FinFET (2) with the same geometric parameters.

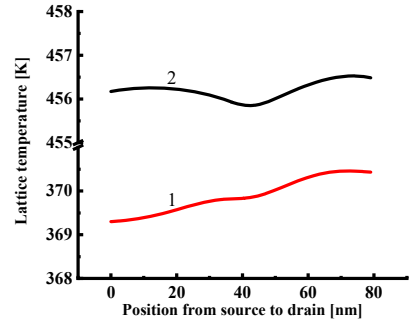

Fig. 4. Distribution of the lattice temperature along the channel of inversion mode FinFET (1) and accumulation mode JL FinFET (2) with the same geometric parameters. Vd=0.75V Vg=1.6V

Drain currents can be influenced by the parameters of the source and drain areas. We simulated the influence of the doping profile and doping depth L (Fig.5) of the source and drain areas to the lattice temperature in the center of the channel.

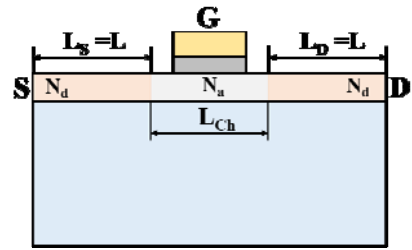

Fig. 5. The structure of the simulated inversion mode FinFET. L is the doping depth of the source and drain areas.

From the simulation results it is seen that with increasing the doping depth of the source and drain areas the lattice temperature is increased for analytical as well as for constant doping profile (Fig.6). With increasing the doping depth up to contacting of the source area with the drain area and when the channel will become homogeneous the lattice temperature of the inversion mode FinFET aspires to the temperature of the JL FinFET.

In Fig.6 we can also see, that the lattice temperature sufficiently depends on the doping profile. In the case of a constant doping profile, the lattice temperature in the channel center is higher by 75-85 K than in the analytical doping profile, depending on the doping depth. This difference is explained by the difference in the current density. In Fig.7 it is seen, that the current density in the channel center in the case of the constant doping profile is higher than in the case of the analytical doping profile. The doping level of the channel in JL FinFET ($5 \cdot 10^{18}$ cm$^{-3}$) is higher than in FinFET by 3 orders and therefore drain current density is higher respectively.

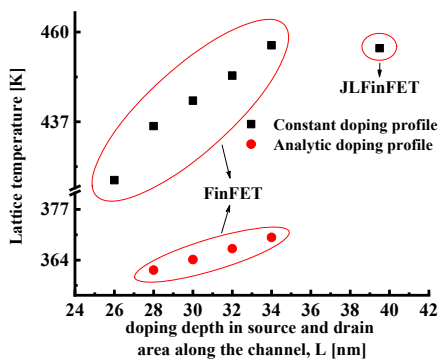

Fig.6. Dependence of the lattice temperature in the center of the channel on the doping depth along the channel.

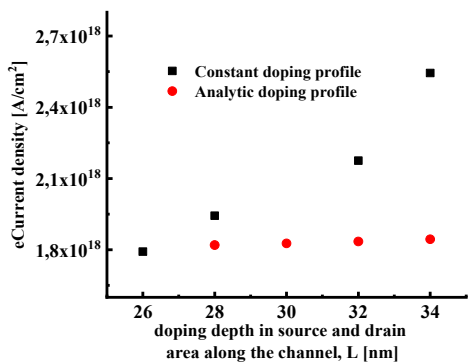

Fig.7. Dependence of the current density in the center of the channel on the doping depth along the channel in FinFET.

## IV. CONCLUSION

Thus SHE is more significant in SOI JL FinFET than in SOI FinFET with the same geometrical parameters. It is explained by differences in the structures and in some electrical parameters. Particularly it is connected with a relatively high doping level in the channel (on several orders) in JL FinFET.

The lattice temperature in the channel center of the inversion mode FinFET depends on the doping profile and doping depth of the source and drain areas. At the constant doping profile the temperature is higher than at the analytical doping profile. With increasing the doping depth along the channel the temperature aspires to the value which corresponds to the temperature of JL FinFET with the same geometry.